\documentclass[conference,10pt]{IEEEtran}

\usepackage{amsfonts}
\usepackage{epsfig} 
\usepackage{amsmath} 
\usepackage{amssymb}  
\usepackage{graphicx}

\newtheorem{theorem}{Theorem}[section]
\newtheorem{lemma}[theorem]{Lemma}
\newtheorem{definition}[theorem]{Definition}

\newcommand{\sr}{\stackrel}

\newcommand{\rar}{\rightarrow}

\newcommand{\tri}{\sr{\triangle}{=}}

\newcommand{\noi}{\noindent}

\newcommand{\be}{\begin{equation}}
\newcommand{\ee}{\end{equation}}
\newcommand{\bea}{\begin{eqnarray}}
\newcommand{\eea}{\end{eqnarray}}
\newcommand{\bes}{\begin{eqnarray*}}
\newcommand{\ees}{\end{eqnarray*}}

\newcommand{\bfi}{\begin{figure}}
\newcommand{\bfit}{\begin{figure}[t]}
\newcommand{\bfib}{\begin{figure}[b]}
\newcommand{\bfih}{\begin{figure}[h]}
\newcommand{\bfip}{\begin{figure}[p]}
\newcommand{\efi}{\end{figure}}

\newcommand{\bi}{\begin{itemize}}
\newcommand{\ei}{\end{itemize}}
\newcommand{\ben}{\begin{enumerate}}
\newcommand{\een}{\end{enumerate}}
\numberwithin{equation}{section} 

\begin{document}
%
\title{Realizable Rate Distortion Function and Bayesian Filtering Theory}

\author{\IEEEauthorblockN{\bf Photios A. Stavrou, Charalambos D. Charalambous and Christos K. Kourtellaris}
\IEEEauthorblockA{ECE Department, University of Cyprus, Green Park, Aglantzias 91,\\
 P.O. Box 20537, 1687, Nicosia, Cyprus\\
e-mail:\{\it stavrou.fotios, chadcha, kourtellaris.christos\}{\it@ucy.ac.cy}}}


\maketitle

\begin{abstract}
 The relation between rate distortion function (RDF) and Bayesian filtering theory is discussed. The relation is established by imposing a causal or realizability constraint  on the reconstruction conditional distribution of the RDF, leading to the definition of a causal RDF. Existence of the optimal reconstruction distribution of the causal RDF is shown using the topology of weak convergence of probability measures. The optimal non-stationary causal reproduction conditional distribution of the causal RDF is derived in closed form; it is given by a set of recursive equations which are computed backward in time. The realization of causal RDF is described via the source-channel matching approach, while an example is briefly discussed to illustrate the concepts.
\end{abstract}


\IEEEpeerreviewmaketitle

\section{INTRODUCTION}

\noi Shannon's information theory for reliable communication evolved over the years without much emphasis on real-time realizability or causality imposed on the communication sub-systems. In particular, the classical rate distortion function (RDF) for source data compression deals with the characterization of the optimal reconstruction conditional  distribution subject to a fidelity criterion \cite{berger,cover-thomas}, without regard for realizability. Hence, coding schemes which achieve the RDF are not realizable.\\
\noi On the other hand, filtering theory is developed by imposing real-time realizability on estimators with respect to measurement data. Specifically, least-squares filtering theory deals with the characterization of the conditional distribution of the unobserved process given the measurement data, via a stochastic differential equation which causally depends on the observation data.\\
  Although, both reliable communication and filtering (state estimation for control) are concerned with the reconstruction of processes, the main underlying assumptions characterizing them are different. There are, however, examples in which the gap between the two disciplines in both the underlying assumption and the form of reconstruction is bridged \cite{berger,liptser-shiryaev1978,ihara1993,cover-pombra1989,charalambous2008}.
In information theory, the real-time realizability or causality of a communication system is addressed via joint  source-channel coding \cite{gastpar2003} (for memoryless channels and sources).

\noi Historically, the work of R. Bucy  \cite{bucy} appears to be the first to consider the direct relation between distortion rate function and filtering, by carrying out the computation of a realizable distortion rate function with square criteria for two samples of the Ornstein-Uhlenbeck process. The earlier work of A. K. Gorbunov and M. S. Pinsker \cite{gorbunov91} on $\epsilon$-entropy defined via a causal constraint on the reproduction distribution of the RDF, although not directly related to the realizability question pursued by  Bucy, computes the causal RDF for stationary Gaussian processes via power spectral densities. The realizability constraints imposed on the reproduction conditional distribution in \cite{bucy} and \cite{gorbunov91} are different, the actual computation of the distortion rate or RDF in these works is based on the Gaussianity of the process, while no general theory is developed to handle arbitrary processes. \\
\noi The objective of this paper is to  develop the general theory by further investigating  the connection between realizable rate distortion theory and filtering theory for general distortion functions and random processes on abstract Polish spaces. The connection is established  via optimization  on the spaces of conditional distributions which satisfy a causality constraint and an average distortion constraint. \\
The main results obtained are the following.
\begin{enumerate}
\item[a)] Existence of optimal reconstruction distribution minimizing the causal RDF  using the topology of weak convergence of probability measures on Polish spaces.
\item[b)] Closed form expression of the optimal reconstruction conditional distribution for non-stationary processes, via recursive equations computed backward in time.
\item[c)] Realization procedure of the filter based on the causal RDF.
\item[d)] Example to demonstrate the realization of the filter.
\end{enumerate}
Although, the operational meaning of the  causal RDF in terms of causal and sequential codes is not pursued, it is pointed out that by utilizing the assumptions and coding theorem derived in \cite{tatikonda2000}, the causal RDF derived is the optimal performance theoretically achievable (OPTA) for sequential codes, while it is related to the OPTA for causal codes \cite{neuhoff1982}.\\
\noi Next, we give a high level discussion on RDF and filtering theory, and discuss their connection.\\
Consider a discrete-time process $X^n\tri\{X_0,X_1,\ldots,X_n\}\in{\cal X}_{0,n} \tri \times_{i=0}^n{\cal X}_i$, and its reconstruction $Y^n\tri\{Y_0,Y_1,\ldots,Y_n\}\in{\cal Y}_{0,n} \tri \times_{i=0}^n{\cal Y}_i$ where ${\cal X}_i$ and ${\cal Y}_i$ are Polish spaces.

\noi{\it Bayesian Estimation Theory.} In classical filtering, one is given a mathematical model that generates the process $X^n$, $\{P_{X_i|X^{i-1}}(dx_i|x^{i-1}):i=0,1,\ldots,n\}$, often induced via discrete-time recursive dynamics, a mathematical model that generates observed data obtained from sensors, say, $Z^n$, $\{P_{Z_i|Z^{i-1},X^i}$ $(dz_i|z^{i-1},x^i):i=0,1,\ldots,n\}$, while $Y^n$ are the causal estimates of some function of the process $X^n$ based on the observed data $Z^n$.
The classical Kalman Filter is a well-known example \cite{kalman1960}, where $\widehat{X}_i =\mathbb{E}[X_i | Z^{i-1}],~i=0,1,\ldots,n$, is the conditional mean which minimizes the average least-squares estimation error. Thus, in classical filtering theory both  models which generate the unobserved and observed processes, $X^n$ and $Z^n$, respectively, are given \'a priori. Fig. 1 is the block diagram of the filtering problem.	
\begin{figure}[ht]
\centering
\includegraphics[scale=0.60]{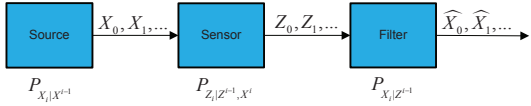}
\caption{Block Diagram of Filtering Problem}
\label{filtering}
\end{figure}

\noi{\it Causal Rate Distortion Theory and Estimation.} In causal rate distortion theory one is given a  distribution for the process $X^n$, which induces $\{P_{X_i|X^{i-1}}(dx_i|x^{i-1}):~i=0,1,\ldots,n\}$, and determines the causal reconstruction conditional distribution $\{P_{Y_i|Y^{i-1},X^i}(dy_i|y^{i-1},x^i):~i=0,1,\ldots,n\}$ which minimizes the mutual information between $X^n$ and $Y^n$ subject to distortion fidelity constraint, via a causal (realizability) constraint. The filter $\{Y_i:~i=0,1,\ldots,n\}$ of $\{X_i:~i=0,1,\ldots,n\}$ is found by realizing the reconstruction distribution $\{P_{Y_i|X^{i-1},X^i}(dy_i|y^{i-1},x^i):~i=0,1,\ldots,n\}$ via a cascade of sub-systems as shown in Fig. 2.
\begin{figure}[ht]
\vspace*{0.1cm}
\centering
\includegraphics[scale=0.60]{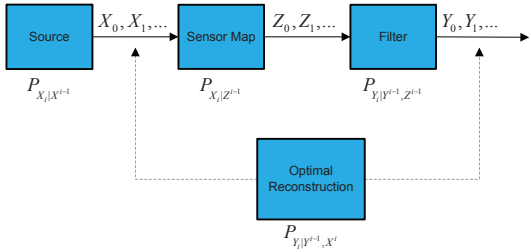}
\caption{Block Diagram of Filtering via Causal Rate Distortion Function}
\label{filtering_and_causal}
\end{figure}

\noi The precise problem formulation necessitates  the definitions of  distortion function or fidelity, and mutual information.\\
The distortion function or fidelity between  $x^n$ and its reconstruction $y^n$, is a measurable function defined by
\begin{align*}
d_{0,n} : {\cal X}_{0,n} \times {\cal Y}_{0,n} \rar [0, \infty], \: \: d_{0,n}(x^n,y^n)\tri\sum^n_{i=0}\rho_{0,i}(x^i,y^i)
\end{align*}
The mutual information between $X^n$ and $Y^n$, for a given distribution ${P}_{X^n}(dx^n)$, and  conditional distribution $P_{Y^n|X^n}(dy^n|x^n)$, is defined by \cite{cover-thomas}
\begin{align}
I(X^n;Y^n)&\tri\int\log\Big(\frac{P_{Y^n|X^n}(dy^n|x^n)}{{P}_{Y^n}(dy^n)}\Big)\nonumber\\
&P_{Y^n|X^n}(dy^n|x^n)\otimes{P}_{X^n}(dx^n) \label{1}
\end{align}
The realizability constraint is introduced next.
\noi Define the causal $(n+1)-$fold convolution measure
\begin{equation}
{\overrightarrow P}_{Y^n|X^n}(dy^n|x^n) \tri \otimes^n_{i=0}P_{Y_i|Y^{i-1},X^i}(dy_i|y^{i-1},x^i)-a.s. \label{9}
\end{equation}
The realizability constraint for a causal filter is defined by
\begin{align}
{\overrightarrow Q}_{ad} \tri&\Big\{ P_{Y^n|X^n}(dy^n|x^n) :\nonumber\\
&P_{Y^n|X^n}(dy^n|x^n) ={\overrightarrow P}_{Y^n|X^n}(dy^n|x^n)-a.s. \Big\}  \label{eq18}
\end{align}
The realizability condition (\ref{eq18}) is necessary, otherwise the connection between filtering and realizable rate distortion theory cannot be established. This is due to the fact that $P_{Y^n|X^n}(dy^n|x^n)=\otimes_{i=0}^n{P}_{Y_i|Y^{i-1},X^n}(dy_i|y^{i-1},x^n)-a.s.$, and hence in general, for each $i=0,1, \ldots,n$, the conditional distribution of $Y_i$ depends on future symbols $\{X_{i+1},X_{i+2},\ldots,X_n\}$ in addition to the past and present symbols $\{Y^{i-1},X^i\}$.\\
\noi {\it Causal RDF.}
The causal RDF is defined by
\begin{equation}
{R}^c_{0,n}(D)\tri \inf_{{ P}_{Y^n|X^n}(dy^n|x^n)\in  \overrightarrow{Q}_{ad}: \mathbb{E}\big\{d_{0,n}(X^n,Y^n)\leq{D}\big\}}I(X^n; Y^n)\label{7}
\end{equation}
Note that realizability condition (\ref{eq18}) is different from the realizability condition  in \cite{bucy}, which is defined under the assumption that $Y_i$ is independent of $X_{j|i}^*\tri X_j -\mathbb{E}\Big(X_j|X^i\Big), j=i+1, i+2, \ldots,$. The claim here is that realizability condition (\ref{eq18}) is more natural and applies to processes which are not necessarily Gaussian having square error distortion function. Realizability condition (\ref{eq18}) is weaker that the causality condition found in \cite{gorbunov91} defined by  $X_{n+1}^\infty \leftrightarrow X^n \leftrightarrow Y^n$.  \\
 The point to be made regarding  (\ref{7}) is that the realizability constraint ${P}_{Y^n|X^n}(dy^n|x^n)= {\overrightarrow P}_{Y^n|X^n}(dy^n|x^n)-a.s.,$ is equivalent to the following (see also Lemma~\ref{lem1}):
\begin{align}
&{P}_{Y^n|X^n}(dy^n|x^n)= {\overrightarrow P}_{Y^n|X^n}(dy^n|x^n)-a.s.
{\Longleftrightarrow} \nonumber\\  &I(X^n;Y^n)=\int \log\Big(\frac{{\overrightarrow P}_{Y^n|X^n}(dy^n|x^n)}{{P}_{Y^n}(dy^n)}\Big)\nonumber\\
&{\overrightarrow P}_{Y^n|X^n}(dy^n|x^n){P}_{X^n}(dx^n)
\equiv{\mathbb I}(P_{X^n},{\overrightarrow P}_{Y^n|X^n})\label{eq6}
\end{align}
where ${\mathbb I}(P_{X^n},{\overrightarrow P}_{Y^n|X^n})$ indicates the functional dependence of $I(X^n;{Y^n})$ on $\{P_{X^n},{\overrightarrow P}_{Y^n|X^n}\}$.\\
Therefore, by finding the solution of  (\ref{7}), then one can realize it via a channel from which one can construct an optimal filter causally as in Fig.~\ref{filtering_and_causal}.\\
This paper is organized as follows. Section~\ref{abstract} discusses the formulation on abstract spaces. Section~\ref{existence} establishes  existence of optimal minimizing  distribution,  and Section~\ref{solution} derives the non-stationary solution recursively. Section~\ref{realization-causal} describes the  realization of causal RDF, while Section~\ref{example} provides an example. Lengthy derivations are omitted due to space limitation.
\section{CAUSAL RDF ON ABSTRACT SPACES}\label{abstract}
 The source and reconstruction alphabets are sequences of Polish spaces \cite{dupuis-ellis97} as defined in the previous section. Probability distributions on any measurable space  $( {\cal Z}, {\cal B}({\cal Z}))$ are denoted by ${\cal M}_1({\cal Z})$.
It is assumed  that the $\sigma$-algebras $\sigma\{X^{-1}\}=\sigma\{Y^{-1}\}=\{\emptyset,\Omega\}$. For $({\cal X}, {\cal B}({\cal X})), ({\cal Y}, {\cal B}({\cal Y}))$  measurable spaces, the set of conditional distributions  $P_{Y|X}(\cdot|X=x)$ is denoted by ${\cal Q}({\cal Y};{\cal X})$ and it is equivalent to stochastic kernels.

\noi Mutual information is defined via the Kullback-Leibler distance:
\begin{align}
&I(X^n;Y^n) \tri  \mathbb{D}(P_{X^n,Y^n}|| P_{X^n}\times{P_{Y^n}})\nonumber\\
& = \int_{{\cal X}_{0,n} \times {\cal Y}_{0,n}} \log \Big( \frac{P_{Y^n|X^n}(dy^n|x^n)}{P_{Y^n}(dy^n)} \Big)P_{Y^n|X^n}(dy^n|x^n) \nonumber\\
&\otimes{P}_{X^n}(dx^n)=\int_{{\cal X}_{0,n}} \mathbb{D}(P_{Y^n|X^n}(\cdot|x^n)|| P_{Y^n}(\cdot))P_{X^n}(dx^n)\nonumber\\
&\equiv \mathbb{I}(P_{X^n}, P_{Y^n|X^n})  \label{re3}
 \end{align}
Note that (\ref{re3}) states that mutual information is expressed as a functional of $\{P_{X^n}, P_{Y^n|X^n}\}$.\\
\noi The next lemma (stated without prove) relates causal product conditional distributions  and conditional independence.
\begin{lemma} \label{lem1}
The following are equivalent.
\begin{enumerate}
\item $P_{Y^n|X^n} (dy^n| x^n)={\overrightarrow P}_{Y^n|X^n}(dy^n|x^n)-a.s.$.

\item For each $i=0,1,\ldots, n-1$,  $Y_i \leftrightarrow (X^i, Y^{i-1}) \leftrightarrow (X_{i+1}, X_{i+2}, \ldots, X_n)$ forms a Markov chain.



\item For each  $i=0,1,\ldots, n-1$, $Y^i \leftrightarrow X^i \leftrightarrow X_{i+1}$ forms a Markov chain.
\end{enumerate}
\end{lemma}
According to Lemma~\ref{lem1}, mutual information subject to causality reduces to
\begin{align}
&I(X^n;Y^n)=\int_{{\cal X}_{0,n} \times {\cal Y}_{0,n}} \log \Big( \frac{ \overrightarrow{P}_{Y^n|X^n}(d y^n|x^n)}{P_{Y^n}(dy^n)} \Big)\nonumber\\
&{\overrightarrow P}_{Y^n|X^n}(dy^n|dx^n)\otimes P_{X^n}(dx^n)
\equiv {\mathbb I}(P_{X^n},\overrightarrow{P}_{Y^n|X^n})  \label{ex11}
\end{align}
where $P_{Y^n}(dy^n) = \int  {\overrightarrow P}_{Y^n|X^n}(dy^n|dx^n)\otimes P_{X^n}(dx^n)$, and (\ref{ex11}) states that $I(X^n;Y^n)$ is a functional of $\{P_{X^n},\overrightarrow{P}_{Y^n|X^n}\}$.
Hence, causal RDF is defined by optimizing ${\mathbb I}(P_{X^n},{P}_{Y^n|X^n})$ over ${P}_{Y^n|X^n}$ subject to the realizability constraint ${P}_{Y^n|X^n}(dy^n|x^n)={\overrightarrow {P}}_{Y^n|X^n}(dy^n|x^n)-a.s.,$ which satisfies a distortion constraint.
\begin{definition}\label{def1}
$($Causal Rate Distortion Function$)$
Suppose $d_{0,n}\tri\sum^n_{i=0}\rho_{0,i}(x^i,y^i)$, where $\rho_{0,i}: {\cal X}_{0,i}  \times {\cal Y}_{0,i}\rightarrow [0, \infty)$, is a sequence of ${\cal B}({\cal X}_{0,i}) \times {\cal B }( {\cal Y}_{0,i})$-measurable distortion functions, and let $\overrightarrow{Q}_{0,n}(D)$ (assuming is non-empty) denotes the average distortion or fidelity constraint defined by
\begin{align}
&{\overrightarrow Q_{{0,n}}(D)}\tri\Big\{{P}_{Y^n|X^n} \in {\cal Q}({\cal Y}_{0,n};{\cal X}_{0,n}) :\nonumber\\
&\ell_{d_{0,n}}({P}_{Y^n|X^n})\tri\int_{{\cal X}_{0,n} \times {\cal Y}_{0,n}}d_{0,n}(x^n,y^n)
{P}_{Y^n|X^n}(dy^n|x^n) \nonumber \\
&  \otimes{P}_{X^n}(dx^n)\leq D\Big\}   \bigcap{\overrightarrow Q}_{ad},~D\geq0\label{eq2}
\end{align}
where ${\overrightarrow{Q}_{ad}}$ is the realizability constraint (\ref{eq18}).
The causal RDF is defined by
\begin{align}
{R}^c_{0,n}(D) \tri  \inf_{{{P}_{Y^n|X^n}\in \overrightarrow{Q}_{0,n}(D)}}{\mathbb I}({P}_{X^n},{P}_{Y^n|X^n})\label{ex12}
\end{align}
\end{definition}
Clearly, ${R}^c_{0,n}(D)$ is characterized by minimizing mutual information or equivalently $\mathbb{I}({P}_{X^n},{P}_{Y^n|X^n})$ over $\overrightarrow{Q}_{0,n}(D)$.

\section{EXISTENCE OF OPTIMAL CAUSAL RECONSTRUCTION}\label{existence}
\par In this section, the existence of the minimizing causal product kernel in (\ref{ex12}) is established  by using the topology of weak convergence of probability measures on Polish spaces. Let $BC({\cal Y}_{0,n})$ denotes the set of bounded continuous real-valued functions on ${\cal Y}_{0,n}$.
The assumptions required are the following.
\begin{enumerate}
\item[1)] ${\cal Y}_{0,n}$ is a compact Polish space, ${\cal X}_{0,n}$ is a Polish space;
\item[2)] for all $h(\cdot){\in}BC({\cal Y}_{0,n})$, the function $(x^{n},y^{n-1})\in{\cal X}_{0,n}\times{\cal Y}_{0,n-1}\mapsto\int_{{\cal Y}_n}h(y)P_{Y|Y^{n-1},X^n}(dy|y^{n-1},x^n)\in\mathbb{R}$ is continuous jointly in the variables $(x^{n},y^{n-1})\in{\cal X}_{0,n}\times{\cal Y}_{0,n-1}$;
\item[3)] $d_{0,n}(x^n,\cdot)$ is continuous on ${\cal Y}_{0,n}$;
\item[4)] the distortion level $D$ is such that there exist sequence $(x^n,y^{n})\in{\cal X}_{0,n}\times{\cal Y}_{0,n}$ satisfying $d_{0,n}(x^n,y^{n})<D$.
\end{enumerate}
Note that since it is assumed that ${\cal Y}_{0,n}$ is a compact Polish space, then ${\cal Q}({\cal Y}_{0,n};{\cal X}_{0,n})$ is weakly compact.
\begin{lemma}\label{compactness2}
Assume that conditions 1), 2) hold.\\
Then
\begin{itemize}
\item[1)] The realizability constraint set ${\overrightarrow Q}_{ad}$ is a closed subset of a weakly compact set ${\cal Q}({\cal Y}_{0,n};{\cal X}_{0,n})$ (hence compact).
\item[2)] Under the additional conditions 3), 4)  the set ${\overrightarrow{Q}}_{0,n}(D)$ is a closed subset of ${\overrightarrow Q}_{ad}$ (hence compact).
\end{itemize}
\end{lemma}
The previous results follow from Prohorov's theorem that relates tighness and weak compactness.
\par The next theorem establishes existence of the minimizing reconstruction kernel for (\ref{ex12}); it follows from Lemma~\ref{compactness2} and the lower semicontinuity of $\mathbb{I}(P_{X^n},\cdot)$ with respect to $P_{Y^n|X^n}$.
\begin{theorem}\label{existence_rd}
Suppose the conditions of Lemma~\ref{compactness2} hold. Then ${R}^c_{0,n}(D)$ has a minimum.
\end{theorem}
\section{NON-STATIONARY OPTIMAL RECONSTRUCTION }\label{solution}
\par In this section the form of the optimal causal product reconstruction kernels is derived under non-stationarity assumption.  The Gateaux differential of the  $(n+1)-$fold convolution product ${\overrightarrow {P}}_{Y^n|X^n}(dy^n|x^n)$ should be varied in each direction of ${P}_{Y_i|Y^{i-1}, X^i}(dy_i|y^{i-1}, x^i), i=0, 1, \ldots, n$.

\begin{theorem} \label{th5}
Suppose ${\mathbb I}_{{P}_{X^n}}(P_{Y_i|Y^{i-1}, X^i}: i=0,1,\ldots,n) \tri {\mathbb I}({P}_{X^n},\overrightarrow{P}_{Y^n|X^n})$ is well defined for every $\overrightarrow{P}_{Y^n|X^n}\in{\cal Q}({\cal Y}_{0,n};{\cal X}_{0,n})$ possibly taking values from the set $[0,\infty].$ Then  $\{P_{Y_i|Y^{i-1}, X^i}: i=0,1,\ldots,n\}  \rightarrow {\mathbb I}_{{P}_{X^n}}(P_{Y_i|Y^{i-1}, X^i}: i=0,1,\ldots,n)       $ is
Gateaux differentiable at every point in ${\cal Q}({\cal Y}_{0,n};{\cal X}_{0,n})$,  and the Gateaux
derivative at the points ${P}_{Y_i|Y^{i-1},X^i}^0$ in each direction $\delta{P_{Y_i|Y^{i-1},X^i}}=P_{Y_i|Y^{i-1},X^i}-{P}_{Y_i|Y^{i-1},X^i}^0$,~$i=0,\ldots,n$, is 
\begin{align}
&\delta{\mathbb I}_{{P}_{X^n}}({P}_{Y_i|Y^{i-1},X^i}^0,{P}_{Y_i|Y^{i-1}, X^i}-{P}_{Y_i|Y^{i-1}, X^i}^0: i=0,\ldots,n)\nonumber\\
&=\sum_{i=0}^n\int_{{\cal X}_{0,i}\times{\cal Y}_{0,i}}\log \Bigg(\frac{{P}_{Y_i|Y^{i-1},X^i}^0}{{P}_{Y_i|Y^{i-1}}^0}\Bigg)\frac{d}{d\epsilon}\overrightarrow{P}_{Y^i|X^i}^{\epsilon}\Big{|}_{\epsilon=0}{P}_{X^i}(dx^i)\nonumber
\end{align}
where $\overrightarrow{P}_{Y^i|X^i}^\epsilon \tri \otimes_{j=0}^i P_{Y_j|Y^{j-1}, X^j}^\epsilon$,  $P_{Y_j|Y^{j-1}, X^j}^\epsilon= P_{Y_j|Y^{j-1}, X^j}^0+ \epsilon \Big( P_{Y_j|Y^{j-1}, X^j}-P_{Y_j|Y^{j-1}, X^j}^0\Big)$, $j=0,1,\ldots, i,~~i=0,1,\ldots, n$,
\begin{align*}
&\frac{d}{d\epsilon}{P}_{Y_0|X^0}^{\epsilon}\Big{|}_{\epsilon=0}=\delta{P}_{{Y_0|X^0}}\nonumber\\
&\frac{d}{d\epsilon}\overrightarrow{P}_{Y^1|X^1}^{\epsilon}\Big{|}_{\epsilon=0}=\delta{P}_{{Y_0|X^0}}\otimes{P}^0_{Y_1|Y^0,X^1}+{P}^0_{{Y_0|X^0}}\otimes\delta{P}_{{Y_1|Y^0,X^1}}\nonumber\\
&\ldots\\
&\frac{d}{d\epsilon}\overrightarrow{P}_{Y^i|X^i}^{\epsilon}\Big{|}_{\epsilon=0}=\delta{P}_{{Y_0|X^0}}\otimes_{j=1}^i{P}^0_{Y_j|Y^{j-1},X^j}+\\
&{P}^0_{{Y_0|X^0}}\delta{P}_{{Y_1|Y^0,X^1}}\otimes_{j=2}^i{P}^0_{{Y_j|Y^{j-1},X^j}}+\ldots+\\
&\otimes_{j=0}^{i-1}{P}^0_{{Y_j|Y^{j-1},X^j}} \otimes \delta{P}_{{Y_i|Y^{i-1},X^i}} ,~i=0,1,\ldots,n.
\end{align*}
\end{theorem}
\par The constrained problem defined by (\ref{ex12}) can be reformulated using Lagrange multipliers as follows (equivalence of constrained and unconstrained problems follows from \cite{dluenberger69}).
\begin{align}
{R}_{0,n}^c(D) &= \inf_{\overrightarrow{P}_{Y^n|X^n}={\otimes_{i=0}^n{P}_{Y_i|Y^{i-1},X^i}}} \Big\{{{\mathbb I}}({P}_{X^n},\overrightarrow{P}_{Y^n|X^n})-\nonumber\\
&s(\ell_{{d}_{0,n}}(\overrightarrow{P}_{Y^n|X^n})-D)\Big\} \label{ex13}
\end{align}
and  $s \in(-\infty,0]$ is the Lagrange multiplier.\\
Note that ${P}_{Y_i|Y^{i-1},X^i}\in{\cal Q}({\cal Y}_i;{\cal Y}_{0,i-1}\times{\cal X}_{0,i})$, therefore, one should introduce another set of Lagrange multipliers to obtain an optimization problem without constraints. This process is involved, hence  we state the main results.\\
\underline{General Recursions for Non-Stationary Optimal Reconstruction}\\
For $k=0,\ldots,n$
\begin{align*}
&g_{n,n}(x^n,y^n) \tri 0, \: \:
g_{n-k,n}(x^{n-k},y^{n-k})  \nonumber \\
& \tri - \int_{{\cal X}_{n-k+1}}  P_{X_{n-k+1}|X^{n-k}}(dx_{n-k+1}|x^{n-k})\\
&  \log\int_{{\cal Y}_{n-k+1}}e^{s\rho_{0,n-k+1}-g_{n-k+1,n}}P_{Y_{n-k+1}|Y^{n-k}}^*(dy_{n-k+1}|y^{n-k})
\end{align*}
 the optimal reconstruction is given by
\begin{align*}
&P_{Y_{n-k}|Y^{n-k-1},X^{n-k}}^*(dy_{n-k}|y^{n-k-1},x^{n-k})=\\
&\frac{e^{s\rho_{0,n-k}-g_{n-k,n}}P_{Y_{n-k+1}|Y^{n-k}}^*(dy_{n-k}|y^{n-k-1})}{\int_{{\cal Y}_n}e^{s\rho_{0,n-k}-g_{n-k,n}}P_{Y_{n-k}|Y^{n-k-1}}^*(dy_{n-k}|y^{n-k-1})}
\end{align*}
The causal RDF is given by
\begin{align*}
&R_{0,n}^c(D)=sD+\sum_{i=0}^n\int_{{\cal X}_{0,i-1}\times{\cal Y}_{0,i-1}}\\
&\bigg(\otimes_{j=0}^{i-1}P_{X_j|X^{j-1}}(dx_j|x^{j-1})\otimes{P}_{Y_j|Y^{j-1},X^j}^*(dy_j|y^{j-1},x^j)\bigg)\\
&\int_{{\cal X}_i}P_{X_i|X^{i-1}}(dx_i|x^{i-1})\bigg(-\int_{{\cal Y}_i}g_{i,n}P_{Y_i|Y^{i-1},X^i}^*(dy_i|y^{i-1},x^i)\\
&-\log\int_{{\cal Y}_i}e^{s\rho_{0,i}-g_{i,n}}P_{Y_i|Y^{i-1}}^*(dy_i|y^{i-1})\bigg)
\end{align*}

\noi The above recursions illustrate the causality, since $g_{n-k,n}(x^{n-k}, y^{n-k})$ appearing in the exponent of the reconstruction distribution integrate out future reconstruction distributions. Note also that for the stationary case  all reconstruction conditional distributions are the same and hence, $g_{n-k,n}(\cdot, \cdot)=0, k=0, 1, \ldots, n$. The above recursions  are general, while depending on the application they can be simplified considerably.

\section{REALIZATION OF CAUSAL RDF}\label{realization-causal}
\begin{figure}[ht]
\centering
\includegraphics[scale=0.60]{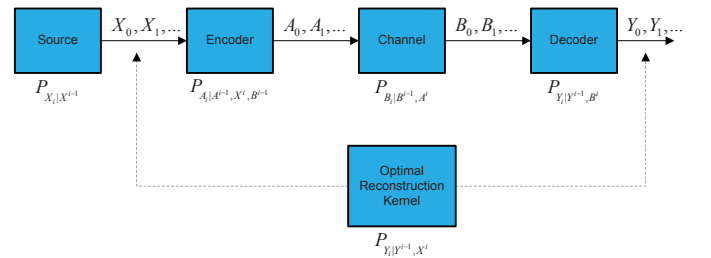}
\caption{Block Diagram of Realizable Causal Rate Distortion Function}
\label{realization_figure}
\end{figure}

The realization of the causal RDF (optimal reconstruction kernel) is equivalent to identifying a communication channel, an encoder and a decoder such that the reconstruction from the sequence $X^n$ to the sequence $Y^n$ matches the causal rate distortion minimizing reconstruction kernel. Fig.~\ref{realization_figure} illustrates the cascade  sub-systems that realize the causal RDF.
This is called source-channel matching in information theory \cite{gastpar2003}. It is also described in \cite{charalambous2008} and \cite{tatikonda2000}; this technique allows one to design encoding/decoding schemes without encoding and decoding delays. The realization of the optimal reconstruction kernel is given below.
\begin{definition}\label{realization}
Given a source $\{P_{X_i|X^{i-1},Y^{i-1}}(dx_i|x^{i-1},y^{i-1}):i=0,\ldots,n\}$,  a channel $\{P_{B_i|B^{i-1},A^{i}}(db_i|b^{i-1},a^i):i=0,\ldots,n\}$ is a realization of the optimal reconstruction distribution if there exists a pre-channel encoder $\{P_{A_i|A^{i-1},B^{i-1},X^i}(da_i|a^{i-1},b^{i-1},x^i):i=0,\ldots,n\}$ and a post-channel decoder $\{P_{Y_i|Y^{i-1},B^i}(dy_i|y^{i-1},b^i):i=0,\ldots,n\}$ such that
\begin{align}
 {\overrightarrow {P}}_{Y^{n}|X^{n}}^*(dy^n|x^n) &\tri\otimes_{i=0}^n P^*_{Y_i|Y^{i-1},X^i}(dy_i|y^{i-1},x^i)-a.s.\nonumber
\end{align}
where the joint distribution is
\begin{align}
&P_{X^n, A^n, B^n, Y^n}(dx^n,da^n,db^n,dy^n) \nonumber \\
&=\otimes_{i=0}^n P_{Y_i|Y^{i-1},B^i}(dy_i|y^{i-1},b^i)  \otimes P_{B_i|B^{i-1},A^{i}}
(db_i|b^{i-1},a^i) \nonumber \\
& \otimes P_{A_i|A^{i-1},B^{i-1},X^i}(da_i|a^{i-1},b^{i-1},x^i) \nonumber \\
 & \otimes P_{X_i|X^{i-1},Y^{i-1}}(dx_i|x^{i-1},y^{i-1}) \nonumber
\end{align}
\end{definition}
The filter is given by $\{P_{X_i|B^{i-1}}(dx_i|b^{i-1}):i=0,\ldots,n\}$.
\noi Thus, if $\{P_{B_i|B^{i-1},A^{i}}(db_i|b^{i-1},a^i):i=0,\ldots,n\}$ is a realization of the causal RDF minimizing distribution  then the channel connecting the source, encoder, channel, decoder achieves the causal RDF, and the filter is obtained.

\section{EXAMPLE: BINARY MARKOV SOURCE}\label{example}
Consider  a binary Markov source, while  the objective is to detect consecutive sequences of $\{1\}$'s subject to a specific, pre-defined distortion or error criterion.  The Markov source  has the following transition probability matrix.
\begin{align}
&P(x_i=0|x_{i-1}=0)=1-p,  \ P(x_i=1|x_{i-1}=0)=p \nonumber \\
&P(x_i=0;|x_{i-1}=1)=q,  \ P(x_i=1|x_{i-1}=1)=1-q \nonumber
\end{align}
The steady state joint probabilities $P(x_i, x_{i-1})$ are given by
\begin{align}
& P(x_i=0, x_{i-1}=0)=\frac{(1-p)q}{p+q} \nonumber \\
& P(x_i=0, x_{i-1}=1)=\frac{pq}{p+q} = P(x_i=1, x_{i-1}=0) \nonumber \\
& P(x_i=1, x_{i-1}=1)=\frac{p(1-q)}{p+q} \nonumber
\end{align}
The distortion function is  described in Table~I.
\vspace*{-0.2cm}
\begin{center}
\begin{table}[h!b!p!]
\begin{center}
$(x_{i},x_{i-1})$
\end{center}
\begin{center}
\begin{tabular}{|l|l|l|l|l|}
\hline  & \ \  00 \ \ & \ \ 01 \ \ & \ \ 10 \ \ & \ \ 11 \ \ \\
\hline $y_i=0$ & \ \ 0 \ \ & \ \ 0\ \  & \ \ 0 \ \ & \ \ 1 \ \ \\
\hline $y_i=1$ & \ \ 1 \ \ & \ \ 1 \ \ & \ \ 1 \ \ & \ \ 0 \ \ \\
\hline
\end{tabular}\label{table1}
\end{center}
\caption{Distortion: $d(x_i,x_{i-1},y_i)$}
\end{table}
\end{center}
\vspace*{-.6cm}
For the given distortion measure the optimal reconstruction kernel
has the following form
\begin{equation}
{P}^{*}(y_i|x_i,x_{i-1})=\frac{e^{sd(x_i,x_{i-1},y_i)}P^*({y}_{i})}{\int_{{\cal  Y}_{i}}
e^{sd(x_i,x_{i-1},y_i)}P^*({y}_{i})} \nonumber
\end{equation}
in which ${P}^{*}(y_i|y_{i-1})=P^*({y}_{i}).$ The Lagrange parameter $s$ is the slope of the causal RDF. Then
\begin{align}
&P^*(1|0,0)=P^*(1|0,1)=P^*(1|1,0)=1- \alpha \nonumber\\
&P^*(0|0,0)=P^*(0|0,1)=P^*(0|1,0)=\alpha \nonumber\\
&P^*(0|1,1)=1-P^*(1|1,1)=1-\beta \nonumber\\
&P^*(y_i=0)=1-P^*(y_i=1)=\gamma \nonumber
\end{align}
where $\alpha=\frac{(1-D)(q-Dp-Dq+pq)}{q(1-2D)(1+p)}$, $\beta=\frac{(1-D)(Dp-p+Dq+pq)}{p(1-2D)(1+q)}$,
$\gamma=\frac{q-Dp-Dq+pq}{(1-2D)(p+q)}$. The causal RDF   is
\[ R^c(D) = \left\{ \begin{array}{ll}
         H\Big(\frac{q(1+p)}{p+q}\Big) -H(D) & \mbox{if $D \leq D_{max}$}\\
        0 & \mbox{if $D > D_{max}$}\end{array} \right. \]
\begin{align}
D_{max}&=\min_{y_i}\sum_{x_i,x_{i-1}}P(d{x}_i,d{x}_{i-1}){d}(x_i,x_{i-1},y_i)\nonumber \\
&=\min\Big(\frac{q(1+p)}{p+q},\frac{p(1-q)}{p+q}\Big) \nonumber
\end{align}
\begin{figure}[ht]
\vspace*{-0.1cm}
\centering
\includegraphics[scale=0.50]{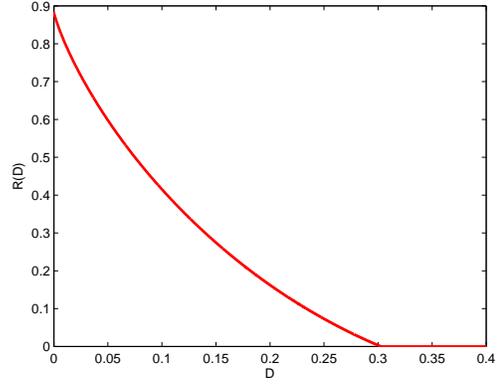}
\caption{$R^c$(D) for p=0.55 and q=0.45}
\label{crdf_for_binary_source}
\end{figure}
\noi The filter which realizes the optimal reproduction kernel $P^*(\cdot|\cdot,\cdot)$ via the specification of an encoder, channel and decoder which achieves the causal RDF, $R^c(D)$, is described in \cite{gastpar2003}.  \\
{\it Special Case.} Consider a special case when  $\frac{q(1+p)}{p+q}=\frac{1}{2}$. Then
\[ R^c(D) = \left\{ \begin{array}{ll}
         1 -H(D) & \mbox{if $D \leq \frac{1}{2}$}\\
        0 & \mbox{if $D > \frac{1}{2}$}\end{array} \right. \]
Note that the capacity of a binary symmetric channel with error probability $\epsilon =D < \frac{1}{2}$ is precisely $C(\epsilon)= 1-H(D)$ \cite{cover-thomas}. Therefore, the realization of the reproduction kernel is given by the cascade of encoder, the binary symmetric channel, and decoder such that the directed information including the encoder but not the decoder operates at the capacity $C(\epsilon)= 1-H(D)$, and it is equal to the directed information from the source to the decoder output. Utilizing the capacity achieving encoder and decoder for the binary symmetric channel  found by Horstein  in \cite{horstein1963}, the realization is completed.
\bibliographystyle{IEEEtran}

\bibliography{photis_references}

\begin{thebibliography}{10}
\providecommand{\url}[1]{#1}
\csname url@samestyle\endcsname
\providecommand{\newblock}{\relax}
\providecommand{\bibinfo}[2]{#2}
\providecommand{\BIBentrySTDinterwordspacing}{\spaceskip=0pt\relax}
\providecommand{\BIBentryALTinterwordstretchfactor}{4}
\providecommand{\BIBentryALTinterwordspacing}{\spaceskip=\fontdimen2\font plus
\BIBentryALTinterwordstretchfactor\fontdimen3\font minus
  \fontdimen4\font\relax}
\providecommand{\BIBforeignlanguage}[2]{{%
\expandafter\ifx\csname l@#1\endcsname\relax
\typeout{** WARNING: IEEEtran.bst: No hyphenation pattern has been}%
\typeout{** loaded for the language `#1'. Using the pattern for}%
\typeout{** the default language instead.}%
\else
\language=\csname l@#1\endcsname
\fi
#2}}
\providecommand{\BIBdecl}{\relax}
\BIBdecl

\bibitem{berger}
T.~Berger, \emph{Rate Distortion Theory:~A Mathematical Basis for Data
  Compression}.\hskip 1em plus 0.5em minus 0.4em\relax Englewood Cliffs, NJ:
  Prentice-Hall, 1971.

\bibitem{cover-thomas}
T.~M. Cover and J.~A. Thomas, \emph{Elements of Information Theory},
  2nd~ed.\hskip 1em plus 0.5em minus 0.4em\relax John Wiley \& Sons, Inc.,
  Hoboken, New Jersey, 2006.

\bibitem{liptser-shiryaev1978}
R.~S. Liptser and A.~N. Shiryaev, \emph{Statistics of Random
  Processes:~{II.}~{A}pplications}, 2nd~ed.\hskip 1em plus 0.5em minus
  0.4em\relax Springer-Verlag, Berlin, Heidelberg, New York, 2001.

\bibitem{ihara1993}
S.~Ihara, \emph{Information theory - for continuous systems}.\hskip 1em plus
  0.5em minus 0.4em\relax World Scientific, 1993.

\bibitem{cover-pombra1989}
T.~Cover and S.~Pombra, ``{G}aussian feedback capacity,'' \emph{IEEE
  Transactions on Information Theory}, vol.~35, no.~1, pp. 37--43, {J}an. 1989.

\bibitem{charalambous2008}
C.~D. Charalambous and A.~Farhadi, ``{LQG} optimality and separation principle
  for general discrete time partially observed stochastic systems over finite
  capacity communication channels,'' \emph{Automatica}, vol.~44, no.~12, pp.
  3181--3188, 2008.

\bibitem{gastpar2003}
M.~Gastpar, B.~Rimoldi, and M.~Vetterli, ``To code, or not to code: {L}ossy
  source-channel communication revisited,'' \emph{IEEE Transactions on
  Information Theory}, vol.~49, no.~5, pp. 1147--1158, {M}ay 2003.

\bibitem{bucy}
R.~Bucy, ``Distortion rate theory and filtering,'' \emph{IEEE Transactions on
  Information Theory}, vol.~28, no.~2, pp. 336--340, {M}ar. 1982.

\bibitem{gorbunov91}
A.~K. Gorbunov and M.~S. Pinsker, ``Asymptotic behavior of nonanticipative
  epsilon-entropy for {G}aussian processes,'' \emph{Problems of Information
  Transmission}, vol.~27, no.~4, pp. 361--365, 1991.

\bibitem{tatikonda2000}
S.~C. Tatikonda, ``Control over communication constraints,'' Ph.D.
  dissertation, Mass. Inst. of Tech.~(M.I.T.), Cambridge,~MA, 2000.

\bibitem{neuhoff1982}
D.~Neuhoff and R.~Gilbert, ``Causal source codes,'' \emph{IEEE Transactions on
  Information Theory}, vol.~28, no.~5, pp. 701--713, {S}ep. 1982.

\bibitem{kalman1960}
R.~E. Kalman, ``A new approach to linear filtering and prediction problems,''
  \emph{Journal of Basic Engineering on Transactions of the ASME}, vol.~82, no.
  Series D, pp. 35--45, March 1960.

\bibitem{dupuis-ellis97}
P.~Dupuis and R.~S. Ellis, \emph{A Weak Convergence Approach to the Theory of
  Large Deviations}.\hskip 1em plus 0.5em minus 0.4em\relax John Wiley \& Sons,
  Inc., New York, 1997.

\bibitem{dluenberger69}
D.~G. Luenberger, \emph{Optimization by Vector Space Methods}.\hskip 1em plus
  0.5em minus 0.4em\relax John Wiley \& Sons, Inc., New York, 1969.

\bibitem{horstein1963}
M.~Horstein, ``Sequential transmission using noiseless feedback,'' \emph{IEEE
  Transactions on Information Theory}, vol.~9, no.~3, pp. 136--143, {J}uly
  1963.

\end{thebibliography}

\end{document}